\documentclass{emulateapj}
\usepackage{epsfig}

\begin{document}

	\title{Swift observations of the ultraluminous X-ray source Holmberg IX X--1}
	\author{A.~K.~H.~Kong\altaffilmark{1,5}, Y.~J.~Yang\altaffilmark{2}, T.-C. Yen\altaffilmark{1}, H. Feng\altaffilmark{3}, P. Kaaret\altaffilmark{4}}
	\altaffiltext{1}{Institute of Astronomy and Department of Physics, National Tsing Hua University, Hsinchu 30013, Taiwan; akong@phys.nthu.edu.tw}
	\altaffiltext{2}{Astronomical Institute ``Anton Pannekoek'', University of Amsterdam, Science Park 904, 1098 XH, Amsterdam, the Netherlands}
	\altaffiltext{3}{Department of Engineering Physics and Center for Astrophysics, Tsinghua University, Beijing 100084, China}
	\altaffiltext{4}{Department of Physics and Astronomy, University of Iowa, Van Allen Hall, Iowa City, IA 52242, USA}
	\altaffiltext{5}{Golden Jade Fellow of Kenda Foundation, Taiwan}

\newcommand{\chandra}{{\it Chandra}}
\newcommand{\asca}{{\it ASCA}}
\newcommand{\rosat}{{\it ROSAT}}
\newcommand{\sax}{{\it BeppoSAX}}
\newcommand{\xmm}{{\it XMM-Newton}}
\newcommand{\swift}{{\it Swift}}
\newcommand{\ho}{Holmberg IX X-1}
\newcommand{\lum}{\thinspace\hbox{$\hbox{erg}\thinspace\hbox{s}^{-1}$}}
\newcommand{\flux}{\thinspace\hbox{$\hbox{erg}\thinspace\hbox{cm}^{-2}\thinspace\hbox{s}^{-1}$}}

\begin{abstract}
\ho\ is a well-known ultraluminous X-ray source with an X-ray luminosity of $\sim 10^{40}$\lum. The source has been monitored by the X-ray Telescope of \swift\ regularly. Since 2009 April, the source has been in an extended low luminosity state. We utilize the co-added spectra taken at different luminosity states to study the spectral behavior of the source. Simple power-law and multi-color disk blackbody models can be ruled out. 
The best overall fits, however, are provided by a dual thermal model with a cool blackbody and a warm disk blackbody. This suggests that \ho\ may be a $10 M_\odot$ black hole accreting at 7 times above the Eddington limit or a $100 M_\odot$ maximally rotating black hole accreting at the Eddington limit, and we are observing both the inner regions of the accretion disk and outflows from the compact object. 

\end{abstract}

\keywords{accretion, accretion disks -- binaries: close -- stars: individual: Holmberg IX X-1 --- X-rays: binaries}

\section{Introduction}

Ultraluminous X-ray sources (ULXs) are luminous ($L_X > 10^{39}$\lum) nonnuclear X-ray point-like sources in galaxies with apparent X-ray luminosities above the Eddington limit for a typical stellar-mass ($\sim 10 M_\odot$) black hole. The majority of ULXs are believed to be accreting objects in binary systems due to their strong X-ray flux variability observed on timescales of hours to years. Assuming an isotropic X-ray emission, ULX is the best candidate of intermediate-mass black hole with a mass of $\sim 10^2-10^4 M_\odot$ (Makishima et al. 2000; Miller \& Colbert 2004). While ULX may represent a missing link between stellar-mass black hole and supermassive black hole in galactic center, its formation and evolution is not well understood. 

The X-ray spectral properties may provide some hints about the connection between ULXs and Galactic black hole X-ray binaries. In particular, many of the ULXs can be modeled with a multi-color disk (MCD; Mitsuda et al. 1984) plus power-law model which is a popular spectral model for Galactic black hole binaries (see Remillard \& McClintock 2006). In contrast to Galactic black hole binaries, many ULXs have a cool ($\sim 0.1-0.2$ keV) accretion disk suggesting a black hole mass of $> 100 M_\odot$ (e.g. Kaaret et al. 2003; Miller, Fabian \& Miller 2004a; Winter, Mushotzky \& Reynolds 2006). While Galactic black hole binaries have a good correlation between the spectral shape and luminosity, the behaviors of ULXs are more complex (e.g. Winter et al. 2006; Kong et al. 2007; Feng \& Kaaret 2009; Kajava \& Poutanen 2009; Vierdayanti et al. 2010). ULXs seem to require more complicated accretion geometry invoking outflows, corona, massive donors, and super-Eddington accretion flows (e.g. Stobbart et al. 2006; Poutanen et al. 2007; Patruno \& Zampieri 2008; Gladstone et al. 2009). 

It is possible that ULX is a distinct class of systems comparing to Galactic black hole binaries. Apart from a few very bright ULXs (also known as hyperluminous X-ray sources, e.g. Farrell et al. 2009), a $> 100 M_\odot$ black hole may not require based on the observed luminosity. Instead, black holes with a few tens of solar masses may be more likely. Theoretical models involving binary mergers (Belczynski et al. 2004) and low-metallicity massive ($\sim 40-50 M_\odot$) progenitors (Mapelli et al. 2009; Zampieri \& Roberts 2009) are proposed to explain a population of ULXs with $\sim 30-90 M_\odot$ black holes.

Alternatively, ULX could be a typical stellar-mass black hole with geometrically or relativistically beamed emission (King et al. 2001; K\"ording et al. 2002) so that the X-ray luminosity does not exceed the Eddington limit. Furthermore, the stellar-mass black hole may in fact accrete materials at or above Eddington limit via a slim disk (Ebisawa et al. 2003), or a radiation pressure-dominated accretion disk model (Begelman 2002). It is also possible due to combination of both scenarios (King 2008).

\ho\ is a famous ULX located near the galaxy M81 and it is about 2 arcmin from M81's dwarf companion, Holmberg IX. The source was first discovered by the {\it Einstein Observatory} (Fabbiano 1988) and has been observed by all major X-ray observatories throughout the last 20 years (La Parola et al. 2001). Apart from the X-ray flux variability, \ho\ is also one of the first ULXs shown to have a cool ($\sim 0.1-0.2$ keV) accretion disk, leading to a suggestion of an intermediate-mass black hole accretor (Miller et al. 2004a). 
It is proven that monitoring observations can reveal the physical nature of Galactic X-ray binaries by tracking their flux and spectral evolution as well as their correlation (see Remillard \& McClintock 2006). Until now, it has been quite difficult to monitor ULXs due to their distances and crowding location. \swift\ is the first X-ray telescope with reasonable spatial resolution and sensitivity to perform such observations. Recently, Kaaret \& Feng (2009) reported \swift\ monitoring observations of several ULXs including \ho, NGC 5408 X--1, and NGC 4395 X--2.

In this paper, we report a \swift\ monitoring observation of the ULX, \ho\ with a focus on the spectral behaviors. We describe the observations and data reduction method in \S2. The results are present in \S3, and a discussion is in \S4. 

\begin{figure*}
	\centering
	\psfig{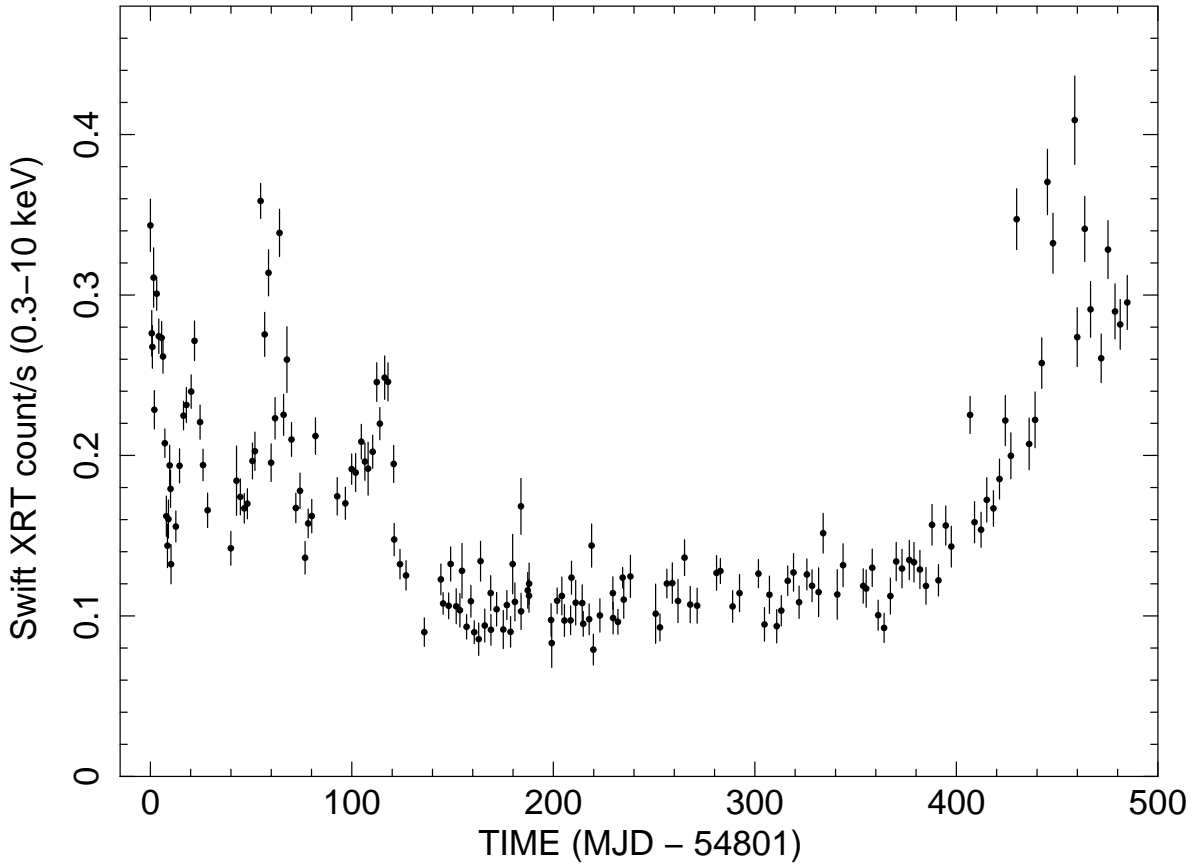}
	\caption{\swift\ XRT 0.3--10 keV lightcurve of \ho\ between 2008 December 1 and 2010 March 31. }
\end{figure*}

\section{Observations and data reduction}

\ho\ is one of the ULXs monitored with the X-ray Telescope (XRT) of 
\swift\ (Kaaret \& Feng 2009). The source has been observed with \swift\ regularly since 2006. In particular, there is a guest observing program (PI: Kaaret) for an intensive monitoring of \ho\ since 2008 December. Furthermore, we proposed a follow-up monitoring program in mid-April 2009 and all the data obtained between April 24, 2009 and 2009 July 23 are from this new program but with the same ObsID (90008). In this work, we focus on the data obtained after 2008 December. In addition, we also include data from ObsID 90079 obtained between 2009 April and 2010 March. We only used data taken in photon counting mode. Timing analysis and hardness variation for data taken between 2006 July and 2009 June have been reported in Kaaret \& Feng (2009).
During the period we are interested, we obtained 177 XRT observations of \ho\ with a total exposure time of 242.1 ks.

We extracted XRT light curves and spectra of \ho\ by using the XRT products generator\footnote{http://www.swift.ac.uk/user\_objects} (Evans et al. 2007, 2009). In brief, this software first creates an image from the event list and identifies our target for each observation. Only events with energy in the range 0.3-10 keV with grades 0--12 are included. A circular source extraction region is chosen to maximize the signal-to-noise ratio of the light curve and spectrum. For the background, an annulus centered on the source is used. Then for each observation, source and background counts are extracted. Source count rates are corrected for the good time interval, and losses due to bad pixels and bad columns. The source spectrum is extracted from a combined event list of all the observations considered. Similarly, the software automatically extracts and combines all the ancillary response files (ARFs); each ARF is weighted according to the proportion of counts in the total source spectrum. Background spectrum is extracted from an annulus region centered on the target excluding any sources in the extraction region. An appropriate redistribution matrix file is selected from the CALDB. 

We grouped the source spectrum with at least 20 counts per spectral bin before fitting in order to allow $\chi^2$ statistics for finding the best-fitting parameters. We performed spectral analysis using the HEAsoft X-ray spectral fitting package XSPEC version 12.5. All spectral fits were made in the 0.3–-10.0 keV band. All errors reported in this work are 90\% confidence errors.

\begin{table*}
	\footnotesize
	\centering
%	\begin{minipage}{120mm}
		\caption{Spectral fit parameters}
		\begin{tabular}{ccccc}
			\hline
			\hline
			Model parameter & Low state & Variable state & High state & All data\\
			\hline
			&& Power-law model\\
			\hline
			$N_H$ ($10^{21}$ cm$^{-2}$) & $1.74\pm0.13$& $2.16\pm0.10$ & $2.96\pm0.23$& $2.16\pm0.07$\\
			$\Gamma$ & $1.74\pm0.04$& $1.86\pm0.03$ & $2.00\pm0.06$& $1.86\pm0.02$\\
			$\chi^2/dof$ & 430.6/315 & 537.5/445 & 238.2/220 & 786.7/535\\
			$L_{0.5-10}$ ($10^{40}$\lum)& 1.00& 1.66 & 2.48& 1.58\\
			\hline
			&& MCD + Power-law model \\
			\hline
			$N_H$ ($10^{21}$ cm$^{-2}$) & $2.60^{+0.64}_{-0.51}$& $2.80^{+0.63}_{-0.51}$ & $4.12^{+1.50}_{-1.28}$ &$3.16^{+0.53}_{-0.44}$\\
			$\Gamma$ & $1.68\pm0.06$& $2.65^{+0.53}_{-0.48}$ & $3.47\pm1.03$ & $3.02^{+0.41}_{-0.37}$\\
			$kT_{in}$ (keV) & $0.20^{+0.06}_{-0.05}$& $2.25^{+0.23}_{-0.18}$ & $1.69^{+0.18}_{-0.15}$ & $2.11^{+0.12}_{-0.11}$\\
			Norm & $71^{+206}_{-55}$& $0.012^{+0.007}_{-0.006}$ & $0.061^{+0.038}_{-0.027}$& $0.016^{+0.005}_{-0.004}$\\
			$\chi^2/dof$ & 410.8/313& 515.7/443 & 205.7/218 &695.4/533\\
			$F_{MCD}/F_{Total}$$^a$ & 0.13 & 0.46 & 0.50 & 0.52\\
			$L_{0.5-10}$ ($10^{40}$\lum)& 1.15& 1.75 & 3.02 & 1.78\\
			\hline
			&& MCD + Blackbody model\\
			\hline
			$N_H$ ($10^{21}$ cm$^{-2}$) & $1.39^{+0.16}_{-0.18}$& $1.20\pm0.10$ & $1.56^{+0.40}_{-0.29}$ &$1.25^{+0.11}_{-0.09}$\\
			$kT$ (keV) & $0.22^{+0.02}_{-0.12}$& $0.26\pm0.01$ & $0.26\pm0.04$& $0.25\pm0.01$\\
			$kT_{in}$ (keV) & $2.05^{+0.12}_{-0.10}$& $1.86\pm0.05$ & $1.60^{+0.10}_{-0.03}$& $1.84\pm0.06$\\
			Norm & $0.014^{+0.001}_{-0.003}$& $0.037\pm0.004$ & $0.086^{+0.028}_{-0.023}$ &$0.033\pm0.004$\\
			$\chi^2/dof$ & 348.2/313& 489.7/443 & 203.7/218 &625.3/533\\
			$F_{MCD}/F_{Total}$$^a$ & 0.85 & 0.85 & 0.86& 0.85\\
			$L_{0.5-10}$ ($10^{40}$\lum)& 0.88& 1.38 & 2.0& 1.35\\
			\hline
			&& DISKPN + EQPAIR model\\
			\hline
			$N_H$ ($10^{21}$ cm$^{-2}$) & $2.22\pm0.25$ & $1.92^{+0.21}_{-0.09}$& $2.26^{+0.38}_{-0.36}$ & $2.01^{+0.10}_{-0.23}$\\
			$T_{max}$ (keV) & $0.26^{+0.06}_{-0.01}$ & $0.28\pm0.02$& $0.31^{+0.11}_{-0.06}$& $0.29\pm0.02$\\
			$l_h/l_s$$^b$ & $5.41^{0.51}_{-0.25}$ & $3.27\pm0.13$ & $3.05^{+1.06}_{-0.88}$& $3.89^{+0.10}_{-0.14}$\\
			$\tau$ & $26.66^{+2.03}_{-1.49}$ & $19.45\pm1.18$ & $23.74\pm6.24$& $23.77^{+0.91}_{-0.99}$\\
			$\chi^2/dof$ & 354.8/312 & 496.7/442 & 203.9/217 & 643.9/532\\
			$L_{0.5-10}$ ($10^{40}$\lum) & 1.03 & 1.72 & 2.17 & 1.47\\
			\hline
			\hline
		\end{tabular}
%	\end{minipage}
\par
\medskip
\begin{minipage}{0.8\linewidth}
	NOTES. --- All quoted errors are 90\% confidence. Luminosities are calculated assuming a distance of 3.6 Mpc.\\
	$^a$ Flux ratio between MCD component and total flux.\\
	$^b$ Ratio between the compactness of electron and the compactness of the seed photon distribution.
\end{minipage}
\end{table*}

\section{Analysis and results}

The X-ray long-term light curve of \ho\ in the 0.3-10 keV is shown in Fig. 1. Our light curve created by the XRT products generator is almost identical to the one generated by Kaaret \& Feng (2009), except that our count rate is systematically slightly higher than theirs. It is likely due to different energy range as well as different correction methods for bad pixels and the point-spread-function. It is very clear from Fig. 1 that the source exhibits substantial variability. In order to search for any modulation in the light curves, we used the Lomb-Scargle periodogram (LSP; Lomb 1976; Scargle 1982), a modification of the discrete Fourier transform which is generalized to the case of uneven spacing. Like Kaaret \& Feng (2009), we found two peaks at $\sim 20$ days and $\sim 60$ days in the LSP. However, both peaks are not statistically significant with a maximum power of 6.6 near a period of 19 days (the 99.9\% significance level has a power of 11.9). The light curve also shows three distinct intensity states. From 2008 December to 2009 early-April, the source count rate varies between 0.15 counts per sec and 0.3 counts per sec, and shows obvious modulation. The source becomes fainter from 2009 mid-April ($\sim$ Day 120 in Fig. 1) to 2010 early-January with an average count rate of 0.1 counts per sec; we defined this period as the ``low'' state. Since 2010 January 10, the source intensity has increased to 0.3--0.4 counts per sec and we name it as the ``high'' state.

Because of the low count rate for each \swift\ observation, we added all the spectra in similar state together to study the spectral behavior. We divided the light curve into three parts: 1) the ``variable'' state with data taken between 2008 December 1 and 2009 April 1; 2) the ``low'' state with data taken between 2009 April 23 and 2010 January 3, and 3) the ``high'' state with data taken between 2010 January 10 and 2010 March 31. We also considered a co-added spectrum from all the data. We first fitted all the spectra with an absorbed power-law model and the spectral parameters are listed in Table 1. All spectra cannot be fitted satisfactorily with a power-law model. Similarly, 
an absorbed MCD blackbody model is not an acceptable fit to any of the spectra when it is the only continuum component. 

We next considered to apply a MCD plus power-law model that provides good fits to a sample of ULXs (e.g. Kaaret et al. 2003; Miller, Fabian \& Miller 2004a; Winter, Mushotzky \& Reynolds 2006). The additional MCD component is statistically significant for all states. For the ``low'' state spectrum, the addition of a disk component is significant at the $> 4\sigma$ level of confidence. In particular, the best-fitting disk temperature is very low with $kT=0.19$ keV (see Table 1), which is consistent with previous observations (Miller et al. 2004b).  
For the ``variable'' and ``high'' state spectra, although the additional MCD component is statistically required, the spectral parameters are completely different comparing to the ``low'' state spectrum. The disk temperature is much higher (1.69--2.25 keV against 0.19 keV) and the photon index is also very different (2.65--3.47 against 1.68). Furthermore, this model is only marginally acceptable for the variable states.

We also fit a dual thermal model consisting of a cool blackbody continuum at low energies and a hot disk blackbody component at high energies (Stobbart et al. 2006). This model provides the best fitting for all states as well as all the data combined with very similar spectral parameters ($kT=0.2$ keV and $kT_{in}=2$ keV; Table 1 and Figure 2). 

Finally we consider a more physical self-consistent Comptonization spectrum using the DISKPN+EQPAIR model (Stobbart et al. 2006; Gladstone et al. 2009). The EQPAIR model (Coppi 1999) allows thermal and non-thermal electron distributions. We tie the temperature of the seed photons to that of the inner accretion disk described by the DISKPN model. This model can describe the ``high'' state data 
equally well as the dual thermal model with a disk temperature similar to that of the cool blackbody component of the dual thermal model. For the ``low'' and ``variable'' states, the fits are slightly worse than that of the dual thermal model.

\begin{figure*}
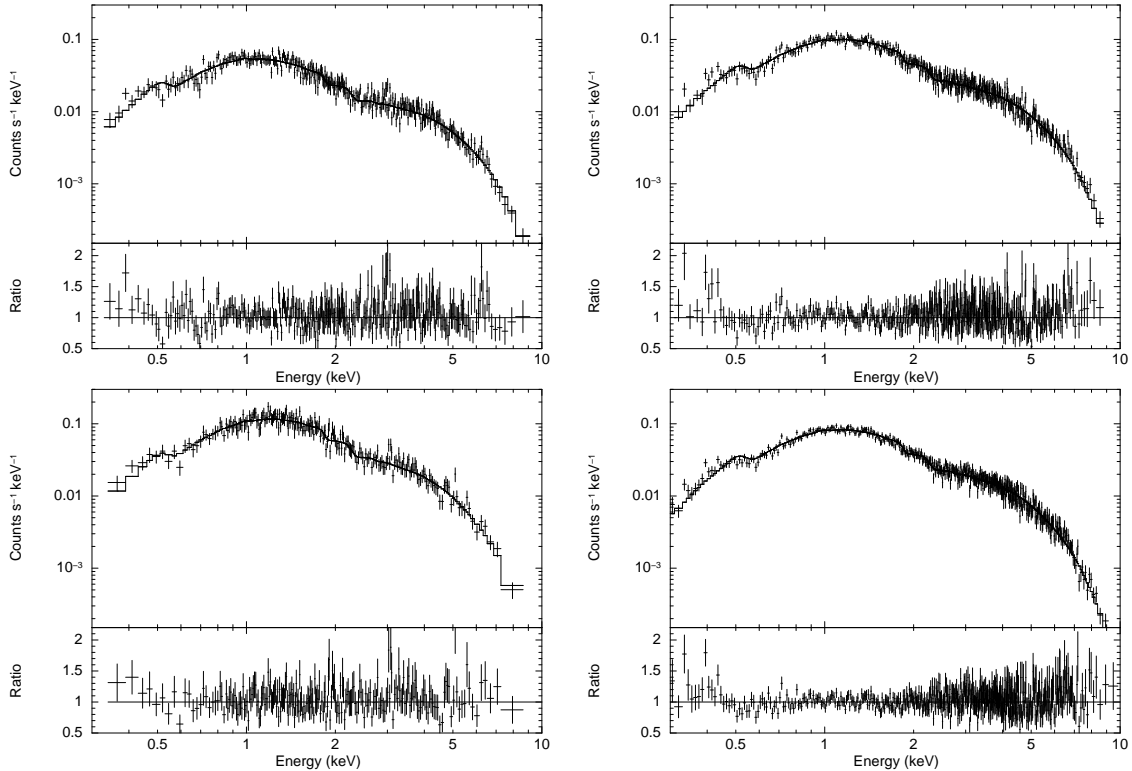

	\centering
	\psfig{file=low.ps,width=2in,angle=-90}
	\hspace*{1.5mm}
	\psfig{file=variable.ps,width=2in,angle=-90}
	\psfig{file=high.ps,width=2in,angle=-90}
	\hspace*{1.5mm}
	\psfig{file=all.ps,width=2in,angle=-90}
	\caption{\swift\ XRT 0.3--10 keV spectra of \ho\ taken during the ``low'' state (upper left), ``variable'' state (upper right), ``high'' state (lower left), and all data (lower right). All spectra can be well described by a dual thermal model with a blackbody temperature of $\sim 0.2$ keV and a MCD temperature of $\sim 2$ keV (See Table 1 for spectral parameters).}
\end{figure*}

\section{Discussion}

We obtained a long-term X-ray light curve of \ho\ by using \swift\ XRT and found that the source transited from a ``variable'' state to a ``low'' state, and then back to a ``high'' state. The co-added spectra of all states can be marginally described with a MCD plus power-law model, and a dual thermal model (blackbody plus MCD) provides the best fits. For the MCD plus power-law model, the ``low'' state spectrum can be best fitted with a cool accretion disk ($kT_{in}=0.19$ keV) plus a power-law ($\Gamma=1.7$). However, during the ``variable'' and ``high'' states, the spectral parameters are completely different (see Table 1). This is very unusual given that the flux difference is only about a factor of 2--3. When we examined the hardness ratios, we did not find any significant change throughout the whole monitoring observation. Kaaret and Feng (2009) also showed by using hardness ratios that the source did not exhibit significant spectral change. We therefore conclude that the apparent spectral change based on spectral fitting may not be real, and it is simply a swap between the MCD and power-law components. More specifically, one common ULX model consists of a cool MCD and a hard power-law just like the spectrum in the ``low'' state; during the ``variable'' and ``high'' states, we have a soft (steep) power-law and a hard MCD. In fact, when we fixed the MCD temperature of the ``variable'' and ``high'' state spectra at the ``low'' state value (0.19 keV), the other spectral parameters are consistent with the ``low'' state. Moreover, Stobbart et al. (2006) rejected the soft power-law plus hard MCD model using \xmm\ data. Thus we do not consider there is a spectral change during the \swift\ monitoring campaign.

While a MCD plus power-law model does not give a consistent result in all the states, the dual thermal model fits the data quite well and it is also the best model among all the spectral models we considered (see Table 1 and Figure 2). This dual thermal model is motivated by the presence of optically thick outflowing winds from a black hole accreting at or above the Eddington limit (King \& Pounds 2003). When this happens, the thick
wind develops a photosphere emitting at a certain temperature.
This blackbody temperature may explain the ultrasoft ($\sim 0.1$ keV) X-ray component of some ULXs. More recently, Stobbart et al. (2006) applied this dual thermal model to describe the \xmm\ spectra of a sample of ULXs (\ho\ is one of the sources) and found out that 10 out of 13 ULXs can be fitted satisfactorily, with $kT = 0.15-0.3$ keV and $kT_{in}=0.8-2.2$ keV. This indicates that we may be observing both the accretion disk as well as the wind from the central black hole. In this scenario, these ULXs are simply the extension of stellar-mass black hole accreting at or above the Eddington limit. 

Lastly, the ``high'' state spectrum can also be fit with a cool accretion disk plus Comptonized corona model. The best-fit parameters are consistent with other ULXs observed with \xmm\ (Gladstone et al. 2009). Based on the fitting statistics (see Table 1), we cannot distinguish the dual thermal model and accretion disk plus Comptonized corona model. For the ``low'' and ``variable''  states, the accretion disk plus Comptonized corona model is slightly worse than the dual thermal model and we cannot totally rule out this model.
It is therefore not clear if the source underwent a spectral change. 
Using similar approach and employing a disk plus thermal Comptonization model, Vierdayanti et al. (2010) assert that spectrum evolves with a definite trend in which the corona temperature decreases and its optical depth increases as the source becomes brighter.  Our data do not support this conclusion.  Indeed, Vierdayanti et al. (2010) present no statistical tests of the correlation of their fitted spectral parameters with luminosity to support this conclusion.
However, there may be some subtle spectral differences in the dual thermal model. For instance, while the ultrasoft soft X-ray component is similar, the best-fit $kT_{in}$ of the ``high'' state is slightly lower than that of the other two states. 

\ho\ has been observed with \xmm\ several times and indeed it is one of the first ULXs to test the MCD plus power-law model (Miller et al. 2004b). In Stobbart et al. (2006), the MCD plus power-law model provides a better fit than the dual thermal model although both are statistically acceptable. In our \swift\ monitoring observation, the dual thermal model (and perhaps also the accretion disk plus Comptonized corona model) can always provide a better fit while the MCD plus power-law model is only acceptable in the ``high'' state data. If this is true, \ho\ may be a black hole accreting at or above the Eddington limit, instead of a massive black hole with $> 100 M_\odot$. This model is also supported by a study of \ho\ using \xmm\ and \asca\ data in which the spectra can be described by a slim disk model indicating that the source is accreting near the Eddington limit (Tsunoda et al. 2006). 

In our dual thermal model, the disk component contributes about 80\% of the total emission which is totally different comparing to the cool disk spectra (in which the MCD component contributes only $\sim 20\%$) discussed by Miller et al. (2004b). Therefore it is likely that the source during our \swift\ monitoring as well as the \xmm\ observation taken in 2001 and \asca\ observation taken in 1999 (Tsunoda et al. 2006) is in a disk-dominant state. We re-analyzed the \xmm\ data used in Tsunoda et al. (2006) and found that the spectrum can also be well described by the dual thermal model with $kT=0.32$ keV and $kT_{in}=1.56$ keV; 
this is consistent with the ``high'' state of our \swift\ observations. We also modeled the \xmm\ data with the DISKPN+EQPAIR model and the best-fit parameters ($T_{max}\sim 0.22$ keV; $\tau\sim 20$) are similar to those of the \swift\ ``high'' state. 

The observed spectral parameters of this dual thermal model are roughly consistent with the calculation of Poutanen et al. (2007) in which the outflow (low temperature component) is from the spherization radius. This also applies to the cool disk component of the DISKPN+EQPAIR model. Following Poutanen et al. (2007), a blackbody temperature of 0.2 keV corresponds to an accretion rate of about 16 times of the Eddington rate for a $100 M_\odot$ black hole. For the high temperature component from the inner disk, the predicted temperature is about 1 keV but Poutanen et al. (2007) noted that it can be up to 4 keV for a rotating black hole. For \ho, the disk temperature is about 2 keV. Following Feng \& Kaaret (2005), the fractional Eddington luminosity can be estimated as 

\begin{equation}
\beta\left(\frac{kT}{1.2 \mbox{keV}}\right)^2 \left(\frac{L_X}{1.3\times10^{39} \mbox{ergs s}^{-1}}\right)^{0.5},
\end{equation}

where $\beta$ is determined by the black hole spin with $\beta=1$ for a Schwarzschild black hole and $1/6$ for a maximally rotating Kerr black hole. Given an X-ray luminosity of $\sim 10^{40}$ ergs s$^{-1}$, \ho\ appears to be above the Eddington limit by a factor of 1.3 even for a maximally rotating black hole.
If the source radiates at the Eddington limit, we estimate that the mass of the black hole is about $100 M_\odot$ based on the observed luminosity. If \ho\ is a non-rotating black hole, the fractional Eddington luminosity can be as large as 7 and the disk temperature corresponds to a $\sim 10 M_\odot$ black hole. 
It is therefore reasonable to expect that the black hole mass of \ho\ is between 10 and 100 $M_\odot$. 
It is worth noting that a $\sim 30-90 M_\odot$ black hole can be formed via binary mergers (Belczynski et al. 2004) or a massive progenitor in a low metallicity environment (Mapelli et al. 2009; Zampieri \& Roberts 2009). 

However, as noted by Stobbart et al. (2006), this dual thermal model requires a specific geometry and viewing angle so that both an optically thick outflow and the inner regions of the accretion disk can be seen at the same time. From recent optical observations, \ho\ is surrounded by a shell-like shock ionized nebula (Miller 1995; Wang 2002; Pakull \& Gris\'e 2008) and this is consistent with our interpretation of the X-ray spectra that strong wind puffs from the disk.

In summary, the nature of the compact object of \ho\ is still a mystery. Previous \xmm\ observations suggest that it is an intermediate-mass black hole based on a cool accretion disk model. Timing study shows that it may be a $50-200 M_\odot$ black hole (Dewangan et al. 2006). Our study using the long-term \swift\ monitoring observations indicates that a dual thermal model can always provide the best fit suggesting a $10-100 M_\odot$ black hole accreting at or above the Eddington limit. 
In order to investigate if there is any spectral change and also the nature of the black hole, \ho\ is therefore deserved for further study using monitoring campaign.

\begin{acknowledgements}
We thank the Swift PI, Neil Gehrels, the Swift science team, and the Swift mission operations team for their support of these observations. We thank Jeanette Gladstone for discussion. This work made use of data supplied by the UK Swift Science Data Centre at the University of Leicester. This project is supported by the National Science Council of the Republic of China (Taiwan) through grant NSC96-2112-M-007-037-MY3. A.K.H.K. gratefully acknowledges support
from a Kenda Foundation Golden Jade Fellowship. Y.J.Y. has received funding from the European Community's Seventh Framework Programme (FP7/2007-2013) under grant agreement number ITN 215212 ``Black Hole Universe''. H.F. is supported by the NSFC under grants 10903004 and 10978001 and by the 973 program 2009CB824800.
\end{acknowledgements}


\begin{references}
	
\reference{} Begelman, M.~C. 2002, ApJ, 568, L97
\reference{} Belczynski,  K., Sadowski,  A., \& Rasio,  F.~A. 2004, ApJ, 611, 1068
\reference{} Coppi, P.~S.\ 1999, High Energy 
Processes in Accreting Black Holes, 161, 375
\reference{} Dewangan, G.~C., Griffiths, R.~E., \& Rao, A.~R.\ 2006, ApJ, 641, L125
\reference{} Ebisawa, K., Zycki, P., Kubota, A. Mizuno, T., \& Watarai, K. 2003, ApJ, 597, 780 
\reference{} Evans, P.~A. et al. 2007, A\&A, 469, 379	
\reference{} Evans, P.~A. et al. 2009, MNRAS, 397, 1177
\reference{} Fabbiano, G. 1988, ApJ, 325, 544
\reference{} Farrell, S.~A., Webb, 
N.~A., Barret, D., Godet, O., \& Rodrigues, J.~M.\ 2009, \nat, 460, 73
\reference{} Feng, H. \& Kaaret, P. 2005, ApJ, 633, 1052
\reference{} Feng, H., \& Kaaret, P.\ 2009, \apj, 696, 1712 
\reference{} Gladstone, J.~C., Roberts, T.~P., \& Done, C.\ 2009, \mnras, 397, 1836 
\reference{} Kaaret, P., Corbel, S., Prestwich, A.~H., \& Zezas, A. 2003, Science, 299, 365
\reference{} Kaaret, P., \& Feng, H. 2009, ApJ, 702, 1679
\reference{} Kajava, J.~J.~E., \& Poutanen, J.\ 2009, \mnras, 398, 1450
\reference{} King, A.~R. Davies, M.~B., Ward, M.~J., Fabbiano, G., \& Elvis, M. 2001, ApJ, 552, L109
\reference{} King, A.~R., \& Pounds, K.~A. 2003, MNRAS, 345, 657
\reference{} King, A.~R.\ 2008, \mnras, 385, L113
\reference{} Kong, A.~K.~H., Yang, 
Y.~J., Hsieh, P.-Y., Mak, D.~S.~Y., \& Pun, C.~S.~J.\ 2007, \apj, 671, 349
\reference{} K\"ording, E., Falcke, H., \& Markoff, S. 2002, A\&A, 382, L13
\reference{} La Parola, V., Peres, G., Fabbiano, G., Kim, D.~W.,  \& Bocchino, F.\ 2001, ApJ, 556, 47
\reference{} Lomb,  N.~R., 1976, Ap\&SS, 39, 447
\reference{} Makishima, K. et al. 2000, ApJ, 535, 632
\reference{} Mapelli, M., Colpi, M., \& Zampieri, L.\ 2009, \mnras, 395, L71
\reference{} Miller, M.~C., \& Colbert, E.~J.~M. 2004, Int. J. Mod. Phys. D, 13, 1
\reference{} Miller, J.~M., Fabian, A.~C., \& Miller M.~C.\ 2004a, ApJ, 614, L117
\reference{} Miller, J.~M., Fabian, A.~C., \& Miller, M.~C.\ 2004b, ApJ, 607, 931
\reference{} Miller, B.~W. 1995, ApJ, 446, L75
\reference{} Mitsuda, K., et al. 1984, PASJ, 36, 741
\reference{} Patruno, A., \& Zampieri, L.\ 2008, \mnras, 386, 543
\reference{} Pakull, M.~W., \&  Gris\'e, F. 2008, AIP Conference Proceedings, 1010, 303 (arXiv:0803.4345)
\reference{} Remillard, R.~A., \& McClintock, J.~E. 2006, ARA\&A, 44 49
\reference{} Scargle,  J.~D., 1982, ApJ, 263, 835
\reference{} Stobbart, A.-M., Roberts, T.P., \& Wilms, J. 2006, MNRAS, 368, 397
\reference{} Tsunoda, N., Kubota, A., Namiki, M., Sugiho, M., Kawabata, K., \& Makishima, K.\ 2006, PASJ, 58, 1081
\reference{} Vierdayanti, K., 
Done, C., Roberts, T.~P., \& Mineshige, S.\ 2010, \mnras, 403, 1206
\reference{} Wang, Q.~D. 2002, MNRAS, 332, 764
\reference{} Winter, L.~M., Mushotzky, R.~F., \& Reynolds, C.~S.\ 2006, ApJ, 649, 730
\reference{} Zampieri, L., \& Roberts, T.~P.\ 2009, \mnras, 400, 677
\end{references}
\end{document}